\documentclass[11pt,doublespace]{article}
\usepackage{natbib}
\usepackage{lastpage}
\usepackage{enumerate}
\usepackage{graphicx}
\usepackage{fullpage}
\usepackage{fancyhdr}
\usepackage{url}
\usepackage{times}
\usepackage{color}
\usepackage{caption}
\usepackage{cite}
\usepackage{natbib}
\usepackage{lastpage}
\usepackage{enumerate}
\usepackage{fullpage}
\usepackage{fancyhdr}
\usepackage{url}
\usepackage{times}
\usepackage{color}
\usepackage{caption}
\usepackage{graphicx}
\usepackage{caption}
\usepackage{subcaption}
\usepackage{algorithm}
\usepackage{algpseudocode}
\usepackage{algorithmicx}
\usepackage{array}
\usepackage{booktabs}
\usepackage[utf8]{inputenc}
\title{Review-Based Rating Prediction}
\date{}
\author{Tal Hadad
\\Dept. of Information Systems Engineering, Ben-Gurion University
\\E-mail: tah@post.bgu.ac.il}

\begin{document}
\maketitle
\lhead{}
\chead{\includegraphics[scale=0.2]{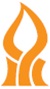} \\ Page \thepage\ of \pageref{LastPage}}
\rhead{}
\setlength{\headheight}{30pt}
\cfoot{Ben-Gurion University of the Negev, Faculty of Engineering Science \\ Department of Information Systems Engineering}
\renewcommand{\headrulewidth}{0.4pt}
\renewcommand{\footrulewidth}{0.4pt}
\begin{abstract}
Recommendation systems are an important units in today's e-commerce applications, such as targeted advertising, personalized marketing and information retrieval.
In recent years, the importance of contextual information has motivated generation of personalized recommendations according to the available contextual information of users.

Compared to the traditional systems which mainly utilize users’ rating history, review-based recommendation hopefully provide more relevant results to users.
We introduce a review-based recommendation approach that obtains contextual information by mining user reviews.
The proposed approach relate to features obtained by analyzing textual reviews using methods developed in Natural Language Processing (NLP) and information retrieval discipline to compute a utility function over a given item.
An item utility is a measure that shows how much it is preferred according to user’s current context.

In our system, the context inference is modeled as similarity between the users reviews history and the item reviews history.
As an example application, we used our method to mine contextual data from customers’ reviews of movies and use it to produce review-based rating prediction.
The predicted ratings can generate recommendations that are item-based and should appear at the recommended items list in the product page.
Our evaluations suggest that our system can help produce better prediction rating scores in comparison to the standard prediction methods.

\end{abstract}

\chapter{Introduction}
In recent years, recommendation systems (RecSys) have been extensively used in various domains to recommend items of interest to users based on their profiles.
RecSys are an integral part of many online stores such as Alibaba.com, Amazon.com, etc.
One of the most famous examples of a recommendation system is Amazon \citep{amazon}. This system contains movie ratings for over 100,000 movies.

A user’s profile is a reflection of the user’s previous selections and preferences that can be captured as rating scores or textual review given to different items in the system.
Using preference data, different systems have been developed to produce personalized recommendations based on collaborative filtering, content-based or a hybrid approach.

Despite the broad used of such recommendation systems, they fail to consider the users' latent preferences, thus may result in performance degradation.
For example, a customer who has once viewed a movie with his friend's child may repeatedly receive suggestions to view kid's movies as the recommendation algorithm select base on the whole history in user's profile without prioritizing his interests.
To address this issue, review-based recommendation systems has been introduced.

Contextual information about a user preference can be explicit or implicit and can be inferred in different ways such as user score ratings or textual reviews.
We concentrate on deriving context from textual reviews.

As an example application of our approach, we have used our method to mine contextual data from customers' reviews of movies domain.

In order to evaluate our method, we have used Amazon movies reviews \citep{amazon}.
The reason for choosing this dataset is that users usually provide some contextual cues in their comments.
For example, they may mention that they are very fond of a specific actor or director, or they may express their opinions about the movie subject or genre that are important to them.
In this dataset each review contains an overall rating, and textual comment.

\chapter{Method}
We assume that user reviews have contextual data about the user preferences, thus comparing the similarity with the item reviews can infer similarity between the two (user preference and item). 
Moreover, similarity between two users' reviews can infer similarity between the two users' preferences. 
We use this approach to predict the rating score that a user will rate an item.
\section{User Context Representation}
\begin{figure}
\centering
\begin{subfigure}{.5\textwidth}
  \centering
  \includegraphics[width=.6\linewidth]{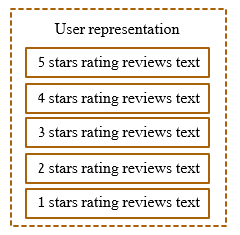}
  \caption{User representation}
  \label{fig:user}
\end{subfigure}%
\begin{subfigure}{.5\textwidth}
  \centering
  \includegraphics[width=.6\linewidth]{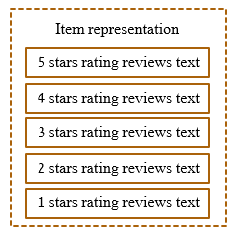}
  \caption{Item representation}
  \label{fig:item}
\end{subfigure}
\caption{Graphic representation of the system two components.}
\label{fig:test}
\end{figure}

Each user will be represented as a set of his reviews, as presented in figure \ref{fig:user}.

User representation composed of 5 strings, for each possible rating value.
Each string is a concatenation of the user's reviews with the corresponding rating value. 
\section{Item Context Representation}
Similar to user representation, each item will be represented as a set of his reviews, as presented in figure \ref{fig:item}.
Item representation composed of 5 strings, for each possible rating value.
Each string is a concatenation of the item's reviews with the corresponding rating value.
\section{Preproccesing}
In this study, we deal with textual dataset (corpuse) which presented in natural-language form (human language), which present difficulties as described in Stanford Handbook \citep{StanfordHandbook}.
Therefore, text normalization is required in order to reduce language diversity including transformation to canonical form for further processing.
We achieve this by performing textual normalization as presented in algorithm~\ref{alg:Textual Normalization}.
The main steps are: (1) removal of punctuation, numbers and stop-words as described in Onix \citep{Onix} ; 
(2) replacement of slang words as described in Twitter Dictionary\citep{TwitterDictionary}; and (3) stem each word  to its root, as described in Stanford Handbook \citep{de2006generating}.
\begin{algorithm}[H]
\caption{Textual Normalization} \label{alg:Textual Normalization}
\textbf{Step 1:} Convert characters into lower cases\\
\textbf{Step 2:} Remove punctuation\\
\textbf{Step 3:} Remove numbers\\
\textbf{Step 4:} Remove stop words, described in Onix \citep{Onix}\\
\textbf{Step 5:} Replace slang terms, described in Twitter Dictionary \citep{TwitterDictionary}\\
\textbf{Step 6:} Word stemming to its root, described in Stanford Handbook \citep{de2006generating}\\
\end{algorithm}
\section{Predicting Item Rating: User-Item Based}
\begin{figure}
\centering
\centering
\includegraphics[width=.6\linewidth]{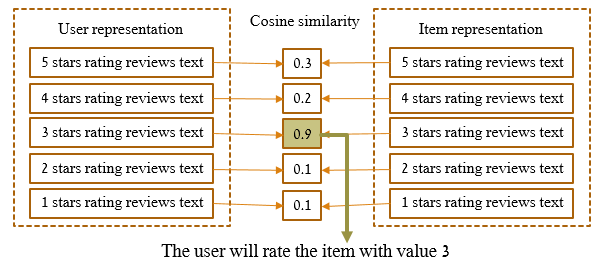}
\caption{Rating prediction process graphic illustration.}
\label{fig:proccess}
\end{figure}

For user $u$ and item $i$, we predict the rating (as presented in figure \ref{fig:proccess}) by computing utility function for each possible rating value $r$ (1-5 stars) as follow:
First, extract and normalize reviews related to $u$ and $i$ (i.e. all reviews written by $u$ or written about $i$).
Second, for each possible rating $r$ ($r\in$\{1,2,3,4,5\}) compare the similarity between $u$ and $i$ reviews who has been rated $r$.
Finally, return the value of $r$ which produced maximum similarity, as the predicted rating for $i$ by $u$.
We present a pseudo code for this process in algorithm \ref{alg:User-Item Rating Prediction Process}, and a graphic representation in figure \ref{fig:proccess}. 
\begin{algorithm}[H]
\caption{User-Item Rating Prediction Process}\label{alg:User-Item Rating Prediction Process}
\begin{algorithmic}[1]
\Procedure{Predict}{$userID,itemID$}
\State $userReviews \leftarrow$ list of $userID$'s reviews
\For{each review $RV_u\in userReviews$}
	\State $RV_u \leftarrow$ normalized text of $RV_u$
\EndFor
\State $itemReviews \leftarrow$ list of $itemID$'s reviews
\For{each review $RV_i\in itemReviews$}
	\State $RV_i \leftarrow$ normalized text of $RV_i$
\EndFor
\For{each possible rating value $r\in$\{1,2,3,4,5\}}
	\State$userReviews_r\leftarrow$reviews whose rated $r$ in $userReviews$
    \State$itemReviews_r\leftarrow$reviews whose rated $r$ in $itemReviews$
    \State$similarity_r\leftarrow$similarity value between $userReviews_r$ and $itemReviews_r$
\EndFor
\State \textbf{return} $\max_{r}(similarity_r)$
\EndProcedure
\end{algorithmic}
\end{algorithm}

Similarity of two textual collections (user and item reviews with same rating score) can be computed in several ways, for that we need to answer two question: (1) \textit{what} to compare, e.g. each user review to each item review or concatenated text of all reviews in each side; and (2) \textit{how} to compare and aggregate, information retrieval discipline have a lot of functions to offer in this subject, e.g. cosine similarity.

\begin{figure}
\centering
\centering
\includegraphics[width=.6\linewidth]{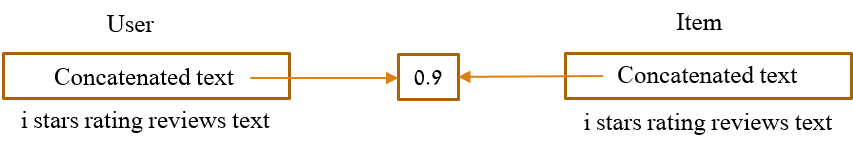}
\caption{CM graphic illustration.}
\label{fig:CM}
\end{figure}

\begin{figure}
\centering
\centering
\includegraphics[width=.6\linewidth]{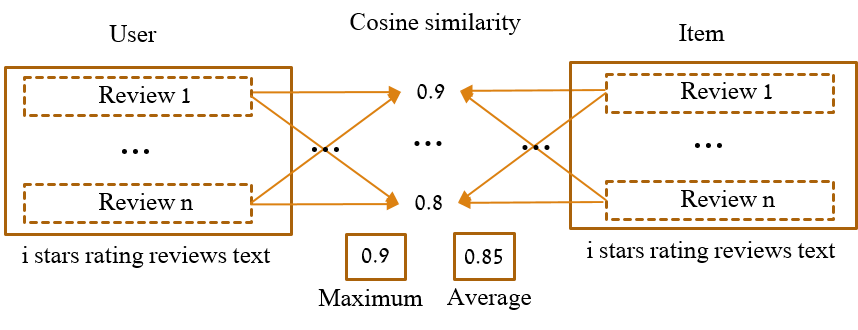}
\caption{ACM and MCM graphic illustration.}
\label{fig:AMCM}
\end{figure}

For user-item based approach we experimented with three different comparison methods, that are different answers to the mentioned questions.
For a given set of user's reviews rated $r$, and a set of item's reviews rated $r$ (line 13 in algorithm \ref{alg:User-Item Rating Prediction Process}), similarity defined in three ways:
\begin{enumerate}
\item $CM$: Cosine similarity between concatenated text of user reviews and concatenated text of item reviews, as presented in figure (\ref{fig:CM}).
\item $MCM$: Maximum cosine similarity between each user review and each item review, as presented in figure (\ref{fig:AMCM}).
\item $ACM$: Average cosine similarity between each user review and each item review, as presented in figure (\ref{fig:AMCM}).
\end{enumerate}

\section{Predicting Item Rating: User-User Based}
Similar User-Item Based approach, for given user $u$ and item $i$, we predict $i$ rating score rated by $u$.
However unlike previous approach, we use collaborate filtering (CF) \citep{ricci2011introduction} method when users similarity ($W$ vector in equation \ref{eq:cf}) defined as cosine similarity between two users textual reviews.
CF is a method of making automatic predictions (filtering) about the interests of a user by collecting preferences or taste information from many users (collaborating).

\begin{figure}
\centering
\centering
\includegraphics[width=.6\linewidth]{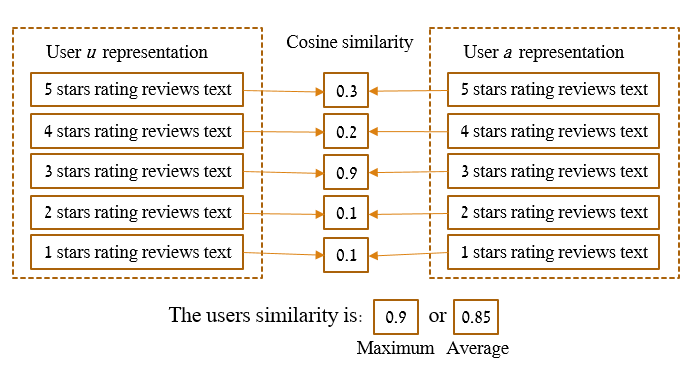}
\caption{CF-MCM and CF-ACM algorithms.}
\label{fig:process2}
\end{figure}

The prediction process is as follow:
First, we extract and normalize reviews that was written by $u$.
Second, for all users $u_i$ that rated item $i$, we extract and normalize reviews that was written by $u_i$.
Third, for each user $u_i$, and for each possible rating $r$ ($r\in$\{1,2,3,4,5\}), we compute cosine similarity between $u$ and $u_i$ reviews that been rated $r$, as presented in figure \ref{fig:process2}.
Finally, using equation \ref{eq:cf}, we predict the rating for item $i$.  

\begin{equation}
\hat{r}_{u,i} = \bar{r_u} + \frac {\sum w_{u_i,u}*(r_{u_i,i}-\bar{r_{u_i}})}{\sum w_{u_i,u}}.
\label{eq:cf}
\end{equation}
Where $\hat{{r}_{u,i}}$ is the predicted item $i$ rating rated by user $u$, $\bar{r_u}$ is ratings average of $u$, $w_{u_i,u}$ similarity value between $u$ and user $u_i$ and $r_{u_i,i}$ rating for item $i$ by $u_i$.

We present a pseudo code for this process in algorithm \ref{alg:User-User Rating Prediction Process}, and a graphic representation of the algorithm in figure \ref{fig:process2}. 

\begin{algorithm}[H]
\caption{User-User Rating Prediction Process}\label{alg:User-User Rating Prediction Process}
\begin{algorithmic}[3]
\Procedure{Predict}{$userID,itemID$}
\State $usersReviews \leftarrow$ Dictionary of Dictionary \{user, \{rating, concatReviews\}\}
\For{each review $RV_u\in usersReviews$}
	\State $RV_u \leftarrow$ normalized text of $RV_u$
\EndFor
\For{each user $u_i\in usersRatings[$itemID$]$}
	\For{each review $RV_u\in usersRevies[$userID$]$}
    	\For{each possible rating value $r\in$\{1,2,3,4,5\}}
          \State$userReviews_r\leftarrow$reviews whose rated $r$ in $usersReviews[userID]$
          \State$u_iReviews_r\leftarrow$reviews whose rated $r$ in $usersReviews[u_i]$
          \State$similarity_r\leftarrow$similarity value between $userReviews_r$ and $u_iReviews_r$
		\EndFor
        \State $w_{u_i,u}$ = $\max_{r}(similarity_r)$
    \EndFor
\EndFor

\State \textbf{return} $\bar{r_u} + \frac {\sum w_{u_i,u}*(r_{u_i,i}-\bar{r_{u_i}})}{\sum w_{u_i,u}}$
\EndProcedure
\end{algorithmic}
\end{algorithm}

Similar User-Item Based approach, we deal with similarity of two textual collections issue, in order to observe there effect on the results. A graphic representation of the two algorithm in figure \ref{fig:process2}. \\
For a given set of the user's ($u$) reviews rated $r$, and a set of users that rated the given item ($u_i$) reviews rated $r_u$ (line 13 in algorithm \ref{alg:User-User Rating Prediction Process}):
\begin{enumerate}
\item $CF-MCM$: Maximum cosine similarity between concatenated text of $u$'s reviews in rating j and concatenated text of $u_i$ reviews in rating j.
\item $CF-ACM$: Average cosine similarity between concatenated text of $u$'s reviews in rating j and concatenated text of $u_i$'s reviews in rating j. (changing line 13 in algorithm \ref{alg:User-User Rating Prediction Process} to AVG).
\end{enumerate}


\chapter{Evaluation}
\section{Dataset\protect\footnote{dataset description was taken from http://snap.stanford.edu/data/web-Amazon.html}}

This dataset consists of movie reviews from amazon.
The data span a period of more than 10 years, including all $\sim$8 million reviews up to October 2012.
Reviews include product and user information, ratings, and a plaintext review.
The dataset can be found at Stanford Large Network Dataset Collection \citep{StanfordDatasets}.\\

\textbf{Source:} J. McAuley and J. Leskovec. From amateurs to connoisseurs: modeling the evolution of user expertise through online reviews. WWW, 2013.\\ \\
\textbf{Dataset statistics:}\\
\begin{tabular}{ |c|c| } \hline
Number of reviews	& 7,911,684\\ \hline
Number of users	& 889,176\\ \hline
Number of products	& 253,059\\ \hline
Users with $>$ 50 reviews	& 16,341\\ \hline
Median no. of words per review	& 101\\ \hline
Timespan & Aug 1997 - Oct 2012\\ \hline
\end{tabular} \\ \\ \\
\textbf{Data Format:}\\
product/productId: B00006HAXW\\
review/userId: A1RSDE90N6RSZF\\
review/profileName: Joseph M. Kotow\\
review/helpfulness: 9/9\\
review/score: 5.0\\
review/time: 1042502400\\
review/summary: Pittsburgh - Home of the OLDIES\\
review/text: "I have all of the doo wop DVD's and this one is as good or better than the 1st ones. Remember once these performers are gone, we'll never get to see them again. Rhino did an excellent job and if you like or love doo wop and Rock n Roll you'll LOVE this DVD !!".\\
Where:
\begin{itemize}
\item product/productId: asin, e.g. amazon.com/dp/B00006HAXW
\item review/userId: id of the user, e.g. A1RSDE90N6RSZF
\item review/profileName: name of the user
\item review/helpfulness: fraction of users who found the review helpful
\item review/score: rating of the product
\item review/time: time of the review (unix time)
\item review/summary: review summary
\item review/text: text of the review
\end{itemize}

\section{Results}

Commonly, datasets for recommendation system evaluation are sparse dataset, thus a second preprocessing phase has been added to the method in order to prune the ratings matrix by removing all those users and items that have less than 5 ratings.
Moreover, due to time limits we could not execute our methods over the $\sim$8 million reviews, thus we used 100,000 reviews randomly selected from the Amazon dataset.
We used two datasets: (1) imbalanced dataset that contains 55\% 5-star reviews, 21\% 4-star reviews, 10\% 3-stars reviews. 6\% 2-stars reviews and 8\% 1-star reviews; and (2) balanced dataset that contains 33\% 5-star reviews, 31\% 4-star reviews, 15\% 3-stars reviews. 9\% 2-stars reviews and 12\% 1-star reviews.
Each dataset was splitted into two parts: (1) 80\% training dataset and (2) 20\% testing dataset, for evaluation purposes.

In previous chapter, we introduced a review-based recommendation system that produce recommendations for a user based on a utility function that depends both the user’s context and also the predicted rating for that item.
As for recommendations that based on predicted ratings, it is logical to use metrics such as MAE and RMSE that compare the predicted rating with the actual ones.

\subsection{MAE and RMSE}
The mean absolute error (MAE) is a quantity used to measure how close predictions are to the real observations. The MSE is given by:

$$ MAE=\frac {1}{n}\sum _{i=1}^{n}\left|f_i-y_i\right|=\frac {1}{n}\sum _{i=1}^{n}\left|e_{i}\right|.$$
The mean absolute error is an average of the absolute errors $|e_i|=|f_i-y_i|$, where $f_i$ is the prediction and $y_i$ the true value.

The root mean square error (RMSE) is also a frequently used measure of the differences between predicted values and the values actually observed. The RMSD represents the sample standard deviation of the differences between predicted values and observed values.
The RMSE is given by:

$$RMSE =\sqrt \frac {\sum _{t=1}^{n}(\hat y_t-y_t)^2}{n}.$$

The RMSE of predicted values $\hat y_t$ for times t of a dependent variable $y_t$ is computed for n different predictions as the square root of the mean of the squares of the deviations.

\subsection{Preproccesing}
First we evaluated how effective our preproccesing phase described in algorithm \ref{alg:Textual Normalization}, used to normalized the texual content by reduce diversity of human language to canonical form.
For this task, we executed $CM$, $MCM$ and $ACM$ algorithms without preproccesing and compared the results to $CM$, $MCM$ and $ACM$ with preproccesing.
Result presented in table \ref{tab:preproccesing} shows a significant improvement when using our preproccesing phase.

\begin{table}[!htbp]
\centering
\begin{tabular}{*5c}
\toprule
Algorithm &  \multicolumn{2}{c}{Preprocessing} & \multicolumn{2}{c}{No Preprocessing}\\
\midrule
{}   & MAE   & RMSE    & MAE   & RMSE\\
$CM$   &  1.37 & 2.02   & 1.73  & 2.89\\
$MCM$   &  0.94 & 1.51   & 1.25  & 2.3\\
$ACM$   &  0.95  &  1.52   & 1.75  & 2.5\\
\bottomrule
\end{tabular}
\caption{Results for evaluate our preprocessing phase.}
\label{tab:preproccesing}
\end{table}

\subsection{Baseline}
Popular automatic predictions method is collaborative filtering (CF) \citep{ricci2011introduction}.
CF is a method of making automatic predictions (filtering) about the interests of a user by collecting preferences or taste information from many users (collaborating).
The underlying assumption of the collaborative filtering approach is that if a person $A$ has the same ratings  as a person $B$ on an issue, $A$ is more likely to have $B$'s opinion on a different issue $x$ than to have the opinion on $x$ of a person chosen randomly.
CF equation to predict user rating over an item was mentioned in equation \ref{eq:cf}, where vector $W_u$ is users similarities for user $u$, computed by pearson ($CF-Pearson$) and cosine ($CF-Cosine$) similarity functions.

In this study we evaluated our method's prediction over the baseline results of CF and other automatic predictions method such as Base Model and Stereotype.
The results are presented in table \ref{tab:bresults}.

\begin{table}[!htbp]
\centering
\begin{tabular}{*5c}
\toprule
Algorithm &  \multicolumn{2}{c}{imbalanced Dataset} & \multicolumn{2}{c}{Balanced Dataset}\\
\midrule
{}   & MAE   & RMSE    & MAE   & RMSE\\
$CM$	& 1.37 & 2.02 &   1.41     &1.37         \\ \hline
$MCM$	& 0.94 & 1.51&    1.01    & 0.94       \\ \hline
$ACM$	& 0.95 & 1.52&       1.45 & 0.95       \\ \hline
$CF-MCM$	& 0.077 & \textbf{0.254}&    0.1002    &\textbf{0.07}        \\ \hline
$CF-ACM$	& 0.092 & 0.299&  0.28      &0.09        \\ \hline
\bottomrule
$CF-Pearson$	& \textbf{0.0722} & 0.646& \textbf{0.0754}       &    0.0722    \\ \hline
$CF-Cosine$	& 0.0883 & 0.7281& 0.0855       &     0.0883   \\ \hline
$Base Model$	& 0.3975 & 0.7424&  0.2763      &     0.3975   \\ \hline
$Stereotype$	& 0.8968 & 1.1162& 1.0137       &       1.1162 \\ \hline
$Random$	& 0.09 & 0.9402&   0.087     &   0.09     \\ \hline
\bottomrule
\end{tabular}
\caption{Results for evaluate our preprocessing phase.}
\label{tab:bresults}
\end{table}

\chapter{Conclusion}
This study has presented a novel approach for mining context from unstructured text and using it to produce predicted rating scores for a given item and given user.
In our proposed methods, the context inference is modeled in two ways: (1) cosine similarity between user and item textual reviews; and (2) cosine similarity between user and other users textual reviews.
The inferred context is used to define a utility function for all possible rating values for an item, reflecting how much each item rating value reviews is similar to a user rating value reviews.

Five novel prediction method was presented: $CM$, $MCM$, $ACM$, $CF-MCM$ and $CF-ACM$.
As an example application, we have used our methods to mine contextual data from customers’ reviews of movies in Amazon's dataset and used it to produce review-based recommendations. 
Our evaluations suggest that our system can help produce better prediction rating scores in comparison to the standard prediction methods.

\bibliographystyle{plain}
\bibliography{main.bib}

\begin{thebibliography}{}


\bibitem[Ricci at al., 2013]{ricci2011introduction}
Ricci at al. (2013).
\newblock {\em Ricci, F., Rokach, L., \& Shapira, B. (2011). Introduction to recommender systems handbook (pp. 1-35). Springer US.}.

\bibitem[Twitter Dictionary, 2016]{TwitterDictionary}
Twitter Dictionary (2016).
\newblock {\em Twitter Dictionary: A Guide to Understanding Twitter Lingo}.
\url{ http://www.webopedia.com/quick_ref/Twitter_Dictionary_Guide.asp}
\newblock {\em [Online; accessed June 2016]}.



\bibitem[Stanford Large Network Dataset Collection, 2016]{StanfordDatasets}
Stanford Large Network Dataset Collection (2016).
\newblock {\em Stanford Large Network Dataset Collection}.
\newblock {\em https://snap.stanford.edu/data/web-Movies.html}
\newblock {\em [Online; accessed June 2016]}.

\bibitem[Amazon, 2016]{amazon}
Amazon (2016).
\newblock {\em Amazon.com: Movies and TV}.
\url{ https://www.amazon.com/movies-tv-dvd-bluray/b/ref=sd_allcat_mov?ie=UTF8&node=2625373011}
\newblock {\em [Online; accessed June 2016]}.


\bibitem[Onix, 2016]{Onix}
Onix (2016).
\newblock {\em Onix Text Retrieval Toolkit}.
\url{ http://www.lextek.com/manuals/onix/stopwords1.html}
\newblock {\em [Online; accessed June 2016]}.


\bibitem[De Marneffe at al., 2006]{de2006generating}
De Marneffe at al. (2006).
\newblock {\em De Marneffe, M. C., MacCartney, B., \& Manning, C. D. (2006, May). Generating typed dependency parses from phrase structure parses. In Proceedings of LREC (Vol. 6, No. 2006, pp. 449-454).}.

\bibitem[Baron at al., 2003]{StanfordHandbook}
Baron at al. (2003).
\newblock {\em Baron, N. S. (2003). Language of the Internet. The Stanford handbook for language engineers, 59-127.}.

\end{thebibliography}
\end{document}